\documentclass[aps,prl,twocolumn,superscriptaddress,showpacs,preprintnumbers,amsmath,amssymb]{revtex4}

\usepackage{graphicx}
\graphicspath{{fig/}}
\usepackage{dcolumn}
\usepackage{bm}
\usepackage{color}

\begin{document}

\preprint{APS/123-QED}

\title{Negative normal restitution coefficient for nanocluster collisions}
\author{Kuniyasu Saitoh}
\affiliation{%
Yukawa Institute for Theoretical Physics, Kyoto University, Sakyo-ku, Kyoto, Japan
}%
\affiliation{%
Department of Mathematics, University of Leicester, Leicester LE1 7RH, United Kingdom
}%
\author{Anna Bodrova}%
\affiliation{%
Yukawa Institute for Theoretical Physics, Kyoto University, Sakyo-ku, Kyoto, Japan
}%
\affiliation{%
Department of Mathematics, University of Leicester, Leicester LE1 7RH, United Kingdom
}%
\affiliation{%
Department of Physics, Moscow State University, Vorobievy Gory 1, 119899 Moscow, Russia
}%
\author{Hisao Hayakawa}%
\affiliation{%
Yukawa Institute for Theoretical Physics, Kyoto University, Sakyo-ku, Kyoto, Japan
}%
\author{Nikolai V. Brilliantov}%
\affiliation{%
Department of Mathematics, University of Leicester, Leicester LE1 7RH, United Kingdom
}%

\date{\today}

\begin{abstract}
The oblique impacts of nanoclusters is studied by means of Molecular Dynamics and theoretically. In
simulations we explore  two models -- Lennard-Jones clusters and particles with covalently  bonded
atoms. In contrast to the case of macroscopic bodies, the standard definition of the normal restitution
coefficient yields for this coefficient negative values for oblique collisions of nanoclusters. We
explain this effect and propose a proper definition of the restitution coefficient which is always
positive. We develop a theory of an oblique impact based on continuum model of particles.  A
surprisingly good agreement between the macroscopic theory and simulations leads to the conclusion, that
macroscopic concepts of elasticity, bulk viscosity and surface tension remain valid for nanoparticles of
a few hundreds atoms.
\end{abstract}


\maketitle

{\it Introduction.} Inelastic collisions, where a part of mechanical energy of colliding bodies
transforms into heat, are common in nature and industry.  Avalanches, rapid granular flows of sand,
powders or cereals may be mentioned as pertinent examples \cite{ine,dom2008}. Moreover, inelastic
collisions define basic properties of astrophysical objects, like planetary rings, dust clouds, etc. An
important characteristic of such collisions is the so-called normal restitution coefficient $e$.
According to a standard definition, it is equal to the ratio of the normal component of the rebound
speed, $\mathbf{g}^{\, \prime}$ (prime states for the post-collision value), and the impact speed,
$\mathbf{g}$
\begin{equation}
\label{eq:defgen_e} e=-\frac{\mathbf{g}^{\, \prime} \cdot
\mathbf{n}}{\mathbf{g} \cdot \mathbf{n}}\, .
\end{equation}
The unit inter-center vector
$\mathbf{n}=\mathbf{r}_{12}/|\mathbf{r}_{12}|$ at the \emph{collision
instant} ($\mathbf{r}_{12}=\mathbf{r}_{1}-\mathbf{r}_{2}$) specifies
the impact geometry. Since particles bounce in the direction, opposite
to that of the impact, $e$ is positive, $e
>0$,  and since the energy is lost in collisions, $e$ is
smaller than one, that is,  $0 \le e \le 1$. This is a common statement in the majority of mechanical
textbook, where it is also claimed that $e$ is a material constant. Recent experimental and theoretical
studies show, however, that the concept of a restitution coefficient is more complicated: First, it
depends on  an impact speed \cite{esp,bri1,bri2}, second, it can exceed unity for a special case of
oblique collisions  with elastoplastic plate \cite{an}, where the energy of normal motion can increase
at the expense of the energy  of tangential motion \cite{an}. Still, it is believed  that $e \le 1$ for
a true head-on collision.
\begin{figure}
\includegraphics[width = 0.8\columnwidth]{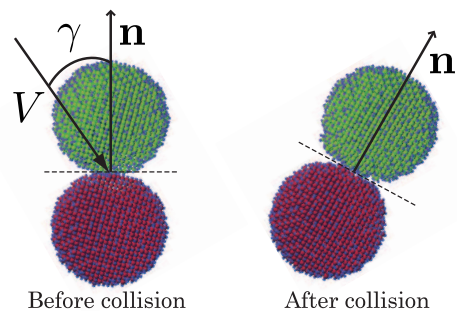}%
\caption{(Color online) Initial (left) and final stage (right) of the nanocluster collision. The initial
relative velocity is $\mathbf{v}_{12}(0)= \mathbf{V}$ and the incident angle is  $\gamma$. The unit
normal $\mathbf{n}$ specifies the orientation of the contact plane. For large $\gamma$ a noticeable
reorientation of this plane is observed. Here the collision of H-passivated Si nanospheres (model B) is
shown. \label{fig:contact}}
\end{figure}
The concept of a restitution coefficient, as a basic one of the classical mechanics,  has been
introduced long ago by Newton; it addresses an impact of macroscopic bodies. The increasing interest to
nanoparticles, inspired by its industrial significance, raises an important question, to what extent the
macroscopic concepts are applicable and whether they acquire new features at a nanoscale.  The
collisions of nanoclusters has been studied in detail numerically \cite{kal,suri,kuni,aws}. It was
observed that the surface effects, due to the direct inter-cluster van der Waals interactions, play a
crucial role: The majority of collisions of homogeneous clusters, built of the same atoms, lead to a
fusion of particles \cite{kal}; they do not fuse for high impact speeds, but disintegrate into pieces
\cite{kal}. This complicates the analysis of restitutive collisions, which may be more easily performed
for particles with a reduced adhesion. Among possible examples of such particles are clusters of
covalently bonded atoms, especially when their surface is coated by atom of different sort, as  for  H-
passivated Si nanospheres \cite{suri}. These particles can rebound from a substrate, keeping their form
after an impact unaltered \cite{suri}. The bouncing nanoclusters demonstrate a surprising effect -- the
normal restitution coefficient can exceed unity even for strictly head-on collisions \cite{kuni}.

In this Letter we investigate the oblique impact of nanoclusters with the reduces adhesion by means of
Molecular Dynamics (MD) and theoretically, using concepts of continuum mechanics. Unexpectedly, we have found
that the normal  restitution coefficient, as defined by Eq.~(\ref{eq:defgen_e}), acquires for large
incident angles negative values, $e<0$. We explain this effect by the reorientation of the contact plane
during an impact and quantify it. Moreover,  we propose a modified definition of $e$, which preserves
its initial physical meaning and yields always positive values. To describe the collision of
nanoclusters  we develop a continuum theory. Surprisingly, the macroscopic approach quantitatively
agrees with MD even at nanoscale.

{\it MD simulations.} We study two models - a simplified model (A), which mimics interactions of
nanoclusters with the reduced adhesion and realistic model (B) for interaction of nanoclusters with
covalently bonded atoms -- H-passivated Si nanospheres. For the model A, which is less computationally
expensive, we adopt the Lennard-Jones (~LJ~) potential $\phi(r) = 4\epsilon \left[ (\sigma_{\rm
LJ}/r)^{12} - (\sigma_{\rm LJ}/r)^6 \right]$ for the interaction between two atoms in the same cluster
and the modified LJ potential $\phi_{\mathrm{int}}(r) = 4\epsilon \left[ (\sigma_{\rm LJ}/r)^{12} - c
(\sigma_{\rm LJ}/r)^6 \right]$ for the interaction between atoms in two different clusters. Here the
cohesive parameter $c=0.2$ controls the adhesive force \cite{kuni,aws} between clusters, while
$\epsilon$, $\sigma_{\rm LJ}$, and $r$ are, respectively,  the depth of potential well, the diameter of
the repulsive core, and the distance between two atoms. The nanoclusters of $N=500$ atoms were prepared
by the two step temperature quench to obtain a rigid amorphous particle \cite{ppt}. The diameter of
nanocluster $d$ was defined as the maximum distance between the center of mass of the nanocluster and
the atom on the surface, so we find $d=10.46\, \sigma_{\rm LJ}$. For the model B we adopt the Tersoff
potential \cite{ter} for the covalent Si-Si, Si-H, and H-H bondings. The Si nanospheres, containing
$2905$ Si atoms arranged in a diamond structure, are  fully coated by $852$ H atoms. The radius of Si
nanosphere is  about $d=2.4 \, \mathrm{nm}$.

\begin{figure}
\includegraphics[width = 0.7\columnwidth]{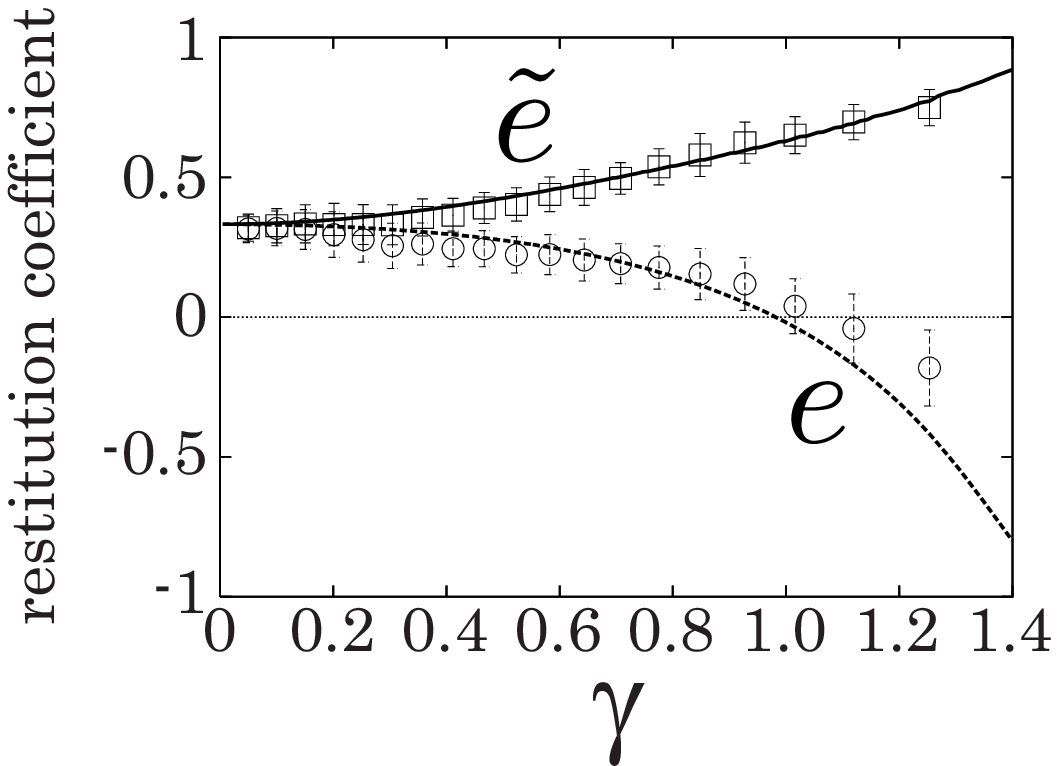}\\
\includegraphics[width = 0.7\columnwidth]{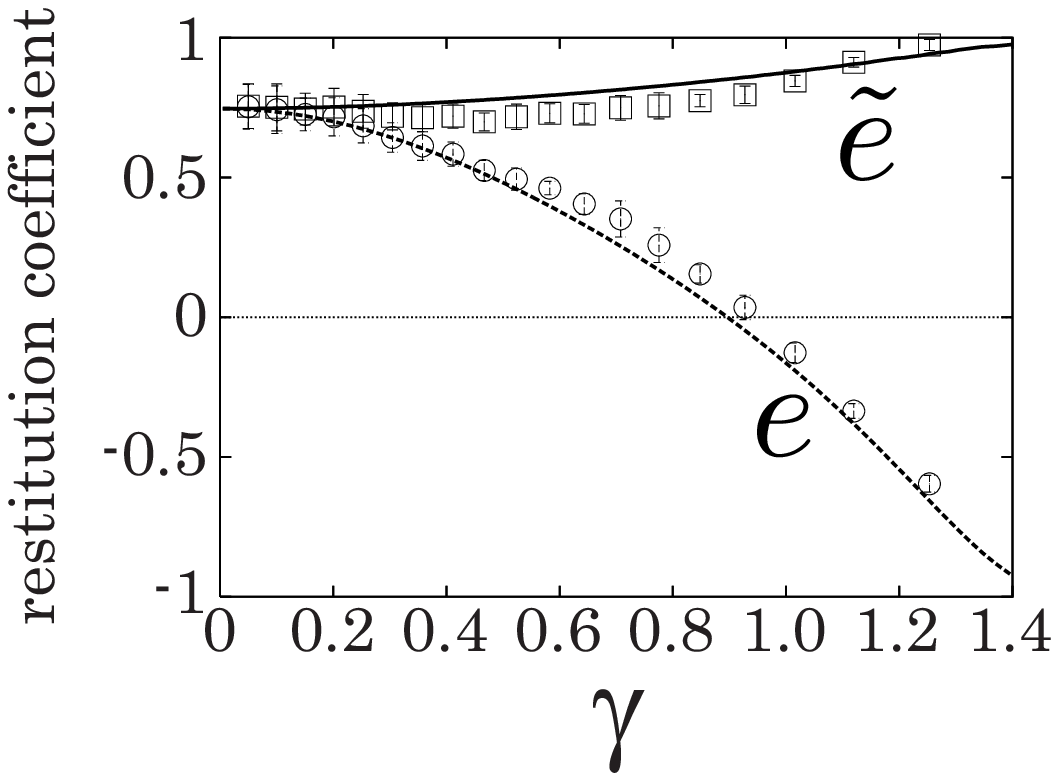}
\caption{ Dependence on the incident angle $\gamma$ of the normal restitution coefficients $e$, and
$\Tilde{e}$ according to the standard definitions (\ref{eq:e_g}) and modified definition (\ref{eq:e'}).
Open circles and squares are respectively the MD results for  $e$ and $\Tilde{e}$, while dashed and
solid lines correspond to theoretical predictions. Upper panel refers to the model A and lower panel --
to the model B. Note that the coefficient $\Tilde{e}$ is always positive. \label{fig:inert}}
\end{figure}
We fix the modulus of the relative inter-cluster velocity
$\mathbf{v}_{1}(0)-\mathbf{v}_{2}(0)=\mathbf{v}_{12}(0)=\mathbf{V}$ and set it  to $V=1.0\,
\sqrt{\epsilon/m}$ and $1850 \, \mathrm{m/s}$ for the model A and B, respectively. We vary the incident
angle $\gamma$ between $\mathbf{n}$ and $\mathbf{V}$ (see Fig. \ref{fig:contact}), so that the
normal impact velocity, $V_n=V \cos \gamma$ is changed. The nanoclusters do not rotate before an impact
and have zero angular velocities, $\mathbf{\omega}_1(0)=\mathbf{\omega}_2(0)=0$. To make an ensemble
average, we randomly turn one of the clusters around the axis, passing through its center and
perpendicular to the contact plane. Due  to rough atomic surfaces of the clusters, this results in
varying contact configurations at  each impact. Hence, for every incident angle $\gamma$ we perform
averaging over 100 collisions with different contact conditions for  model A and over 10 collisions for
model B. The clusters' deformation during an impact is quantified by the normal displacement, $\xi_n
(t)= d - |\mathbf{r}_{12}(t)| = d - r_{12}(t)$. We define the beginning of a collision at $t=0$ and the
end at $t=t_c$ through the conditions, $\xi_n(0)=\xi_n(t_c)=0$.

\begin{figure}
\includegraphics[width = 6 cm]{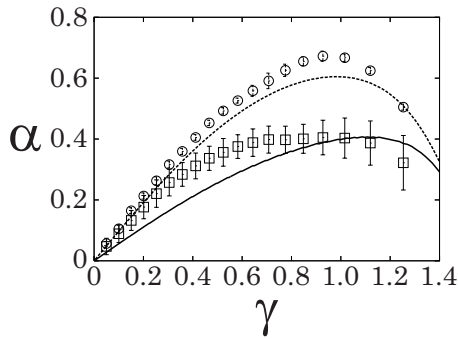}
\caption{ Dependence of the angular displacement $\alpha = {\rm
arcos}\left[\mathbf{n}(0)\cdot\mathbf{n}(t_c)\right]$ of the unit normal  $\mathbf{n}$,
Eq.~(\ref{eq:alpha}), on the incident angle $\gamma$. Open squares and circles are the MD results for
the model A and B, respectively. Solid and broken lines are the corresponding theoretical predictions,
$\alpha = \int_0^{t_c} \Omega (t) dt$ (see text for detail). \label{fig:alpha}}
\end{figure}

Simulation results for the normal restitution coefficient for the models A and B are shown in Fig.
\ref{fig:inert} (upper and lower panel respectively). As it is seen from the figure, the restitution
coefficient $e$,  defined  by Eq.~(\ref{eq:defgen_e}) becomes negative for large incident angles
$\gamma$. Such unusual behavior of $e$ at nanoscales may be understood if we notice that the orientation
of the contact plane, characterized by the unit vector  $\mathbf{n}(t)=\mathbf{r}_{12}(t)/ r_{12}(t)$,
significantly alters during the collision, Fig. \ref{fig:contact}. This is quantified by the angle
$\alpha$ between the initial and final orientations of $\mathbf{n}(t)$,
\begin{equation}
\cos\alpha = \mathbf{n}(0)\cdot\mathbf{n}(t_c)\, .   \label{eq:alpha}
\end{equation}
The dependence of $\alpha$ on the incident angle $\gamma$ is shown in Fig. \ref{fig:alpha}. If $\alpha$
is large, the normal restitution coefficients becomes negative, Fig. \ref{fig:inert}.

\emph{Modified definition of \,$e$:} To analyze this effect, consider the relative velocity of particles
at their contact,
\begin{equation}
\mathbf{g} = \mathbf{v}_{12} + \frac{d}{2} \left[\mathbf{n} \times
\mathbf{\omega}_{12}\right]=-\dot{\xi}_n \mathbf{n} +
r_{12}\dot{\mathbf{n}} + \frac{d}{2} \left[\mathbf{n} \times
\mathbf{\omega}_{12}\right]\, , \label{eq:G}
\end{equation}
where $\mathbf{\omega}_{12} \equiv \mathbf{\omega}_1 + \mathbf{\omega}_2$ and we use $
\mathbf{v}_{12}=\dot{\mathbf{r}}_{12}$ with $\mathbf{r}_{12} = \mathbf{n}(d-\xi_n)$.  In the standard
definition of $e$ and  theoretical studies of an oblique impact \cite{p08}, $\mathbf{n}$ is taken at the
collision instant,  that is, its reorientation during the impact is ignored. In experiments, the normal
$\mathbf{n}$ is also determined only once, at the beginning of an impact  \cite{ex},
$\mathbf{n}=\mathbf{n}(0)$. Neglecting angular velocities (note that $\mathbf{\omega}_{1/2}(0)=0$) we
find for the restitution coefficient:

\begin{equation} e=-\frac{\mathbf{g}(t_c) \cdot
\mathbf{n}(0)}{\mathbf{g}(0) \cdot \mathbf{n}(0)}= \left|
\frac{\dot{\xi}_n(t_c)}{\dot{\xi}_n(0)} \right| \cos \alpha - \frac{ d
\sin \alpha \, \dot{\alpha} }{V \cos \gamma }\, ,  \label{eq:e_g}
\end{equation}
where we take into account that $\mathbf{g}(0) \cdot \mathbf{n}(0) = -\dot{\xi}_n(0)= - V \cos \gamma
<0$, that $r_{12}(t_c)=d$ and  that $\dot{\xi}_n(t_c)<0$. For head-on collisions, when $\gamma \to 0$
and $\alpha \to 0$ (see Fig. \ref{fig:alpha}) the second term in the r.h.s. of Eq. (\ref{eq:e_g}) is
negligible and $e$ is positive. For oblique impacts  $\gamma$ and $\alpha$ are large and   the second
term prevails, yielding a negative $e$.

Hence, the negative restitution coefficient is a consequence of a significant reorientation of a contact
plane during a collision. For hard particles with a small collision duration the reorientation of
$\mathbf{n}$ is small and may be neglected \cite{p08}; this usually holds true for macroscopic bodies.
Nanoclusters, however, are very soft particles with small Young's modulus \cite{mod}. The duration of
their impact $t_c$ is relatively large and the reorientation of the contact plane is significant.

As it follows from the Eq. (\ref{eq:e_g}),  the  standard definition of $e$ characterizes not only the
normal motion along $\mathbf{n}(t)$ (the first term in the r.h.s. of Eq. (\ref{eq:e_g})), but also the
change of the normal $\mathbf{n}(t)$ (the second term in the r.h.s. of Eq. (\ref{eq:e_g})). Therefore,
it is  worth to define the restitution coefficient, which describes pure normal motion. The respective
modification of the standard definition reads:
\begin{equation}
\Tilde{e}=-
\frac{\mathbf{g}(t_c)\cdot\mathbf{n}(t_c)}{\mathbf{g}(0)\cdot\mathbf{n}(0)}
=\left| \frac{\dot{\xi}_n(t_c)}{\dot{\xi}_n(0)} \right|~. \label{eq:e'}
\end{equation}
Here we use Eq. (\ref{eq:G}) for $t=t_c$ and take into account that $\dot{\mathbf{n}}\cdot \mathbf{n}=0$
for a unit vector $\mathbf{n}$. Note, that the modified restitution coefficient $\Tilde{e}$ is always
positive, Fig. \ref{fig:inert}. It can be also seen from Fig. \ref{fig:inert} that the magnitude of
$\Tilde{e}$ for an oblique impact (for large $\gamma$) is significantly larger than that for a head-on
collision. In what follows we explain the observed behaviors of $e$ and $\Tilde{e}$ using a simple
theoretical model, based on continuum mechanics approach.

\emph{Theory of an oblique impact.} Consider a non-inertial frame, rotating with the angular velocity
$\mathbf{\Omega}$, perpendicular $\mathbf{n}$, so that $\dot{\mathbf{n}} =\mathbf{\Omega} \times
\mathbf{n}$. To compute the normal force acting  between two nanoclusters we apply the impact theory
for macroscopic viscoelastic adhesive spheres  \cite{bri1,bri2}. It contains the JKR force \cite{jkr},
which accounts for elastic interactions via the Herzian force $F_H$ and for adhesive interactions via
the Boussinesq force $F_B$,
\begin{equation}
F_H - F_B = \frac{4a^3}{Dd}-\sqrt{\frac{6\pi\sigma}{D}}a^{3/2}~.
\label{eq:FHFB}
\end{equation}
It also contains the dissipative force \cite{bri2},
\begin{equation}
F_D = \dot{a}\eta\left( \frac{12a^2}{Dd} -
\frac{3}{2}\sqrt{\frac{6\pi\sigma}{D}}a^{1/2} \right)~. \label{eq:FD}
\end{equation}
Here, $a$ is the contact radius of the  colliding nanoclusters, related to the normal displacement
$\xi_n$ as
\begin{equation}
\xi_n = \frac{4a^2}{d} - \sqrt{\frac{8\pi\sigma Da}{3}} \, ,
\label{eq:xia}
\end{equation}
and $D=(3/2)(1-\nu^2)/Y$ is the elastic constant with the Young modulus $Y$ and the Poisson ratio $\nu$.
From the independent numerical simulations we estimate $Y=88.3 \, \epsilon/ \sigma_{LJ}^3$ and $\nu
=0.396\,$ for model A, and $Y=283 \,  \mathrm{GPa}$ and $\nu = 0.166$ for model B \cite{mod}.  The
surface tension $\sigma$ may be expressed via Hamaker constant $A_H$ and the equilibrium distance
between atoms at the interface $z_0$ as $\sigma \simeq A_H / 24\pi z_0^2$.  We obtain $\sigma = 0.0246
\, \epsilon/ \sigma_{LJ}^2$ and $0.00289 \, \mathrm{N/m}$ for the models A and B, respectively.  The
dissipative constant $\eta$, which accounts for the viscoelasticity of the particles' material
\cite{bri1} is used here as a fitting parameter. In the present simulations a  good agreement is
obtained by choosing $\eta=0.95\,\sigma_{LJ}\sqrt{m/\epsilon}$  and $1.62 \, \mathrm{fs}$ for models A
and B, respectively.

In the non-inertial frame, the inertial force must be also taken  into account.  Its normal component
reads \cite{Landau:Mechanics},
\begin{equation}
F_I =2 \mu \mathbf{v}_{12} \cdot \dot{\mathbf{n}}(t) - \mu x_{12} |\dot{\mathbf{n}}(t)|^2~,
\label{eq:FI}
\end{equation}
where $\mu=Nm/2$ is the reduced mass of the nanoclusters. If  we again neglect the angular velocities of
particles in the collision (since $\mathbf{\omega}_{1/2}(0)=0$), that is, if we assume that the two
clusters at a contact move together as a solid dumbbell, we can exploit the conservation of the angular
momentum in the form,
\begin{equation}
\mu r_{12}^2  \Omega  = \mu V \sin\gamma \,d \, ,
\end{equation}
where we take into account  that $\mathbf{n} \cdot \mathbf{\Omega}=0$. This yields $\Omega (t) =  V
\sin\gamma \,d/r_{12}^2(t) $ and the inertial force,
\begin{equation}
F_I = \frac{\mu V^2d^2}{r_{12}^3}\sin^2\gamma~. \label{eq:F'}
\end{equation}
Combining  Eqs.~(\ref{eq:FHFB}) -- (\ref{eq:F'}) we obtain the equation
of motion for $\xi_n$:
\begin{equation} \mu \frac{d^{\prime\, 2} }{dt^2}\xi_n + F_H-F_B+F_D +
\frac{\mu V^2d^2}{(d-\xi_n)^3}\sin^2\gamma = 0~, \label{eq:motion}
\end{equation}
where $d^{\prime} /dt$ denotes the time derivative in the non-inertial frame. Solving
Eq.~(\ref{eq:motion})  for $\xi_n(t)$, we obtain $\Tilde{e}$ as it follows from Eq.~(\ref{eq:e'}).
Taking into account that $\dot{\alpha}= \Omega(t_c)$  we obtain from Eq. (\ref{eq:e_g}) the relation
between the standard and modified restitution coefficients,
\begin{equation}
e= \Tilde{e} \cos \alpha - \tan \gamma \sin \alpha ~. \label{eq:e_e'}
\end{equation}
The last equation together with  the relation $\alpha =\int_0^{t_c} \Omega(t) dt$ may be used to compute
the standard coefficient $e$. The theoretical predictions for the coefficients $e$ and  $\Tilde{e}$  are
shown on the upper and lower panels of Fig. \ref{fig:inert} respectively. The agreement between our
theory, which has only one fitting parameter, and MD simulations is rather good. We find that the
restitution coefficient of H-passivated Si nanospheres is well reproduced by our macroscopic theory for
the incident speed between $20$ $\mathrm{m/s}$ and $2405$ $\mathrm{m/s}$. If, however,  the speed
exceeds $2500$ $\mathrm{m/s}$, the nanospheres melt and fuse upon collisions and the theory fails to
describe the impact.

We wish to stress that our theoretical model, developed for  nanoclusters, may be relevant  for oblique
collisions of macroscopic bodies, provided the re-orientation of the contact plane during the impact is
not negligible. This may happen for soft cohesive particles with a low Young modulus and large collision
time. Relevance of the theory for collisions in wet granular systems is also expected \cite{hren}.

\emph{In conclusion,} we perform a detailed study of the oblique impact of nanoclusters by means of
Molecular Dynamics and theoretically. In simulations we use two models, a simplified one, based on the
Lennard-Jones potential with a cohesive parameter and a realistic model for nanoclusters with covalently
bonded atoms. We detect unexpected behavior of the normal restitution coefficient $e$, which becomes
negative for large incident angles and explain this effect by the reorientation of the contact plane in
the course of collision. We propose a modified definition of the restitution coefficient, $\Tilde{e}$,
which describes only the normal motion of particles,  independently of their relative reorientation, and
is always positive. A simple relation between $e$ and $\Tilde{e}$, that may be helpful for experiments
is reported. We develop a theoretical model for an oblique impact, based on the continuum mechanics
description of colliding  particles, and demonstrate that theoretical predictions agree well with
simulation results. Hence, we conclude that the macroscopic concepts of elasticity, surface tension
and bulk viscosity are well applicable for nano-objects of a few hundreds atoms.

We thank H. Kuninaka, D. Rosato and C. M. Hrenya for fruitful discussions.
This work was supported by the Global COE Program
"The Next Generation of Physics, Spun from Universality \& Emergence"
from the Ministry of Education, Culture, Sports, Science and Technology (MEXT) of Japan,
the Research Fellowship of the Japan Society for the Promotion of Science for Young Scientists (JSPS),
and the Grant-in-Aid of MEXT (Grants No.21015016, 21540384 and 21.1958).

\bibliography{oblique}

\end{document}